\documentclass[conference]{IEEEtran}

\usepackage[pdftex]{graphicx}
\usepackage{caption}
\def\Pr{p}
\def\M{\mathcal{M}}
\def\tasks{\textit{tasks}}
\def\attrib{\textit{attribs}}
\def\F{\mathcal{F}}

\usepackage{algorithmic}
\usepackage{algorithm}
\usepackage{amsmath}
\usepackage{booktabs}
\usepackage{subcaption}
\usepackage{caption}
\usepackage{url}

{\itshape}{\rmfamily}
\IEEEoverridecommandlockouts

\begin{document}
	\renewcommand{\thetable}{\arabic{table}}
	\renewcommand{\figurename}{Fig.}

\title{Malware Task Identification: A Data Driven Approach}

\author{\IEEEauthorblockN{Eric Nunes, Casey Buto, Paulo Shakarian\thanks{\noindent U.S. Provisional Patent 62/182,006.  Contact shak@asu.edu for licensing information.}}
	\IEEEauthorblockA{School of Computing, Informatics and\\Decision Systems Engineering\\
		Arizona State University\\
		Tempe, AZ 85281, USA\\
		Email: \{enunes1, cbuto, shak\} @asu.edu}
	\and
	\IEEEauthorblockN{Christian Lebiere, \\Stefano Bennati, \\Robert Thomson}
	\IEEEauthorblockA{Carnegie Mellon University\\
		Pittsburgh, PA 15218\\
		Email: \{cl@cmu.edu ,\\ \{sbennati, thomsonr\} @andrew.cmu.edu\}}
	\and
	\IEEEauthorblockN{Holger Jaenisch\footnote{John Hopkins University (hjaenis@jhu.edu)}}
	\IEEEauthorblockA{Sentar Inc.\\
		Huntsville, AL 35805\\
		Email: holger.jaenisch@sentar.com}
}

\maketitle

\begin{abstract}
	Identifying the tasks a given piece of malware was designed to perform (e.g. logging keystrokes, recording video, establishing remote access, etc.) is a difficult and time-consuming operation that is largely human-driven in practice.  In this paper, we present an automated method to identify malware tasks.  Using two different malware collections, we explore various circumstances for each - including cases where the training data differs significantly from test; where the malware being evaluated employs packing to thwart analytical techniques; and conditions with sparse training data.  We find that this approach consistently out-performs the current state-of-the art software for malware task identification as well as standard machine learning approaches - often achieving an unbiased F1 score of over 0.9.  In the near future, we look to deploy our approach for use by analysts in an operational cyber-security environment.
\end{abstract}

\section{Introduction}
Identifying the tasks a given piece of malware was designed to perform (e.g. logging keystrokes, recording video, establishing remote access, etc.) is a difficult and time consuming task that is largely human-driven in practice~\cite{sikorski}.  The complexity of this task increases substantially when you consider that malware is constantly evolving, and that how each malware instance is classified may be different based on each cyber-security expert's own particular background.  However, automated solutions are highly attractive for this problem as it can significantly reduce the time it takes to conduct remediation in the aftermath of a cyber-attack.

Earlier work has sought to classify malware by similar ``families'' which has been explored as a supervised classification problem~\cite{Bayer_scalable,Kinable:2011,Kong:2013}.  However, differences over ``ground truth'' for malware families (e.g. Symantec and MacAfee cluster malware into families differently) and the tendency for automated approaches to primarily succeed at ``easy to classify'' samples~\cite{Li07onchallenges,PerdisciU12} are two primary drawbacks of malware family classification.  More recently, there has been work on directly inferring the tasks a malware was designed to perform~\cite{invencia}.  This approach leverages static malware analysis (i.e. analysis of the malware sample conducted without execution, such as decompilation) and a comparison with a crowd-source database of code snippets using a proprietary machine leaning approach.  However, a key shortcoming of the static method is that it is of limited value when the malware authors encrypt part of their code -- as we saw with the infamous Gauss malware~\cite{Gauss}.  This work builds upon recent developments in the application of cognitive models to intelligence analysis tasks~\cite{Lebiere:2013} and our own preliminary studies on applying cognitive models to identify the tasks a piece of malware was designed to perform~\cite{iccm:2015,brims}.  Specifically, the contributions of this paper include,
\vspace{0em}
\begin{itemize}
	\setlength{\itemsep}{-1pt}
	\item{Experimental results illustrating consistent and significant performance improvements (in terms of precision, recall, and F1) of the instance-based cognitive model approach when compared with various standard machine learning approaches (including SVM, logistic regression and random forests) for two different sandboxes and for two different datasets.} 
	\item{Experimental results showing a consistent and significant performance improvement of the instance-based cognitive model and several other machine learning approaches when compared to the current state-of-the-art commercial technology (which is based on static analysis).}
	\item{Experiments where we study cases where the malware samples are mutated, encrypted, and use different carriers - providing key insights into how our approach will cope with operational difficulties.}
	\item{Experimental results illustrating that a cognitively-inspired intermediate step of inferring malware families provides improved performance in the machine learning and rule-based cognitive model (though no significant change to the instance-based cognitive model).}
\end{itemize}

This paper is organized as follows. In Section~\ref{tp} we state the technical preliminaries used in the paper. In Section~\ref{apprSec} we introduce our cognitive-based approaches, describing the algorithms and explaining our selection of parameter settings.  This is followed by a description of the baseline approaches that we studied in our evaluation in Section~\ref{baselineAppr} and a description of the two different dynamic malware sandbox environments we used in Section~\ref{sandbox}.  In Section~\ref{expSec} we present our suite of experimental results which include experiments involving samples discovered by Mandiant, Inc. in their APT1 report~\cite{mandiant} and samples created using the GVDG~\cite{gvdg} tool.  Finally, related work and conclusion are discussed in Section~\ref{rwSec} and Section~\ref{con} respectively.

\section{Technical Preliminaries}
\label{tp}
Throughout this paper, we shall assume that we have a set of malware samples that comprise a historical corpus (which we shall denote $\M$) and each sample $i \in \M$ is associated with a set of tasks (denoted $\tasks(i)$) and a set of attributes (denoted $\attrib(i)$).  Attributes are essentially binary features associated with a piece of malware that we can observe using dynamic and/or static analysis while the tasks - which tell us the higher-level purpose of the malware - must be determined by a human reviewing the results of such analysis.  As $\M$ comprises our historical knowledge, we assume that for each $i \in \M$ both $\tasks(i)$ and $\attrib(i)$ are known.  For a new piece of malware, we assume that we only know the attributes.  We also note that throughout the paper, we will use the notation $|\cdot|$ to denote the size of  a given set. Tables~\ref{exAt} and \ref{exTask} provide an example of the attributes and tasks based on the malware samples from the Mandiant APT1 dataset (created from samples available at \cite{apt1sams}, see also \cite{mandiant}).  A full description of this dataset is presented in Section~\ref{expSec}.

\vspace{-1em}
\begin{table}[h!]
	\caption{\textmd{Attributes extracted through automated malware analysis}}
	\label{exAt}
	\centering
	\renewcommand{\arraystretch}{1.5}
	
	\begin{tabular}{|p{2.3cm}|p{5cm}|} 
		\hline
		{\bf Attribute} &  {\bf Intuition} \\ \hline 
	
		\textsf{usesDLL(}\textit{X}\textsf{)} & Malware uses a library \textit{X}\\ \hline
		\textsf{regAct(}\textit{K}\textsf{)} & Malware conducts an activity in the registry, modifying key \textit{K}.\\ \hline
		\textsf{fileAct(}\textit{X}\textsf{)} & Malware conducts an activity on certain file \textit{X}\\ \hline
		\textsf{proAct} & Malware initiates or terminates a process\\ 
		\hline
	\end{tabular}
	\vspace{-1em}
	
\end{table}

\vspace{-1em}
\begin{table}[h!]
	\caption{\textmd{Sample of malware tasks}}
	\label{exTask}
	\centering
	\renewcommand{\arraystretch}{1.5}

	\begin{tabular}{|p{2.2cm}|p{5cm}|} 
		\hline
		{\bf Task} &  {\bf Intuition} \\ \hline 
		\textsf{beacon} & Beacons back to the adversary's system\\  \hline
		\textsf{enumFiles} & Designed to enumerate files on the target\\ \hline
		\textsf{serviceManip} & Manipulates services running on the target\\ \hline
		\textsf{takeScreenShots} & Takes screen shots\\ \hline
		\textsf{upload} & Designed to upload files from the target\\ 
		\hline
	\end{tabular}
	\vspace{-1em}
	
\end{table}

Throughout the paper, we will also often consider malware families, using the symbol $\F$ to denote the set of all families.  Each malware sample will belong to exactly one malware family, and all malware samples belonging to a given family will have the same set of tasks. Hence, we shall also treat each element of $\F$ as a subset of $\M$.

\section{ACT-R Based Approaches}
\label{apprSec}

We propose two models built using the mechanisms of the ACT-R (Adaptive Control of Thought-Rational) cognitive architecture~\cite{Anderson04anintegrated}.  These models leverage the work on applying this architecture to intelligence analysis problems~\cite{Lebiere:2013}.  In particular, we look to leverage our recently-introduced instance-based (ACTR-IB) and rule-based (ACTR-R) models~\cite{iccm:2015,brims}.  Previous research has argued the ability of instance-based learning in complex dynamic situations making it appropriate for sensemaking~\cite{Gonzalez2003591}. On the other hand the rule-based learning is a more compact representation of associating samples in memory with their respective families.  In this section, we review some of the major concepts of the ACT-R framework that are relevant to these models and provide a description of both approaches. 

We leveraged features of the declarative memory and production system of the ACT-R architecture to complete malware task identification. These systems store and retrieve information that correspond to declarative and procedural knowledge, respectively. Declarative information is the knowledge that a person can attend to, reflect upon, and usually articulate in some way (e.g., by declaring it verbally or by gesture). Conversely, procedural knowledge consists of the skills we display in our behavior, generally without conscious awareness. 

\noindent\textbf{Declarative Knowledge.}  Declarative knowledge is represented formally in terms of \textit{chunks}. Chunks have an explicit type, and consist of an ordered list of slot-value pairs of information.  Chunks are retrieved from declarative memory by an activation process, and chunks are each associated with an \textit{activation strength} which in turn is used to compute an \textit{activation probability}.  In this paper, chunks will typically correspond to a malware family.  In the version of ACTR-IB where we do not leverage families, the chunks correspond with samples in the training data. 

For a given chunk $i$, the activation strength $A_i$ is computed as,
\begin{equation}
\label{actStr}
A_i = B_i + S_i + P_i
\end{equation}
where, $B_i$ is the base-level activation, $S_i$ is the spreading activation, and $P_i$ is the partial matching score.  We describe each of these in more detail as follows.\\
{\bf Base-Level Activation ($B_i$):}
The base-level activation for chunk $i$ reflects the frequency of samples belonging to a particular family in memory . More important, base-level is set to the $log$ of the prior probability (i.e., the fraction of samples associated with the chunk) in ACTR-R; for instance-based (ACTR-IB), we set it to a base level constant $\beta_i$.\\ 
{\bf Spreading Activation ($S_i$):}
The spreading activation for chunk $i$ is based on a strength of association between chunk $i$ and the current test malware sample being considered.  The strength of association is computed differently in both approaches and, in some cognitive model implementations, is weighted (as is done in ACTR-R of this paper).\\
{\bf Partial Matching ($P_i$):}
A partial matching mechanism computes the similarity between two samples.  In this work, it is only relevant to the instance-based approach. Given a test sample $j$, its similarity with a sample $i$ in memory is computed as a product of the mismatch penalty ($mp$, a parameter of the system) and the degree of mismatch $M_{ji}$.  We define the value of $M_{ji}$ to be between $0$ and $-1$; $0$ indicates complete match while $-1$ complete mismatch.

As common with models based on the ACT-R framework, we shall discard chunks whose activation strength is below a certain threshold (denoted $\tau$). All the chunks with activation greater than $\tau$ are denoted as $A_j$. Once the activation strength, $A_i$, is computed for a given chunk, we can then calculate the activation probability, $\Pr_i$.  This is the probability that the cognitive model will recall that chunk and is computed using the Boltzmann(softmax) equation~\cite{Sutton:1998}, which we provide below.
\begin{equation}
\label{retprob}
Pr_i = \dfrac{(e^{\frac{A_i}{s}})}{\sum_j (e^{\frac{A_j}{s}}) }
\end{equation}
Here, $e$ is the base of the natural logarithm and $s$ is momentary noise inducing stochasticity by simulating background neural activation (this is also a parameter of the system).  
\vspace{0em}
\subsection{ACT-R Instance-Based Model}
The instance based model is an iterative learning method that reflects the cognitive process of accumulating experiences (in this case the knowledge base of training samples) and using them to predict the tasks for unseen test samples. Each malware instance is associated with a set of attributes. When a new malware sample is encountered, the activation strength of that sample with each sample in memory is computed using Equation~\ref{actStr}. The spreading activation is a measure of the uniqueness of the attributes between a test sample $i$ and a sample $j$ in memory. To compute the spreading activation we compute the $fan$ for each attribute $a$ ($fan(a)$ finds all instances in memory with the attribute $a$) of the test sample $i$.  The Partial matching is computed as explained above.  The degree of mismatch is computed as the intersection between the attribute vector of the given malware and each sample in memory normalized using the Euclidean distance between the two vectors. The retrieval probability of each sample $j$ in memory with respect to the test sample $i$ is then computed using Equation~\ref{retprob}.  This generates a probability distribution over families.  The tasks are then determined by summing up the probability of the families associated with that task with an appropriately set threshold (we set that threshold at $0.5$, based on rationality).

\algsetup{indent=2em}
\begin{algorithm}[]
	\caption{ACT-R Instance-based Learning}
	\begin{algorithmic}

		\STATE \textbf {INPUT:} New malware sample $i$, historical malware corpus $\M$.
		\STATE \textbf {OUTPUT:} Set of tasks associated with sample $i$.
		
		\FOR {query malware sample $i$}
		
		\FORALL {$j$ in $\M$} 
		
		\STATE $B_j = \beta_j$
		
		\STATE $P_j = mp \times \frac{|\attrib(i)\cap\attrib(j)|}{\sqrt{|\attrib(i)| \times |\attrib(j)|}}$
		\STATE $s_{ij}$ = 0.0
		\FOR {$a \in \attrib(i)$}		
		\IF {$a \in \attrib(j)$}
		\STATE $s_{ij}$ += $log(\frac{|\M|}{|fan(a)}|)$ 
		\ELSE
		\STATE $s_{ij}$ += $log(\frac{1}{|\M|})$
		\ENDIF
		\ENDFOR

		\STATE $S_j$ = $\sum_j\frac{s_{ij}}{|\attrib(i)|}$
		\STATE Calculate $A_j$ as per Equation~\ref{actStr}
	
		\ENDFOR
		
		\STATE Calculate $\Pr_j$ as per Equation~\ref{retprob}
		\STATE $\Pr_f = \sum_{j \in f \textit{ s.t. } A_j \geq \tau} \Pr_j$

		\STATE $t_p = \{t \in T | \Pr_f\geq0.5\}$
		\ENDFOR
		
	\end{algorithmic}
	
\end{algorithm}

\subsection{ACT-R Rule-Based Model}
In this version of ACT-R model we classify samples based on simple rules computed during the training phase. Given a malware training sample with its set of attributes $a$, along with the ground truth value family, we compute pair of conditional probabilities $p(a | f)$ and $p(a | \neg f)$ for an attribute in a piece of malware belonging (or not belonging) to family $f$. These probabilistic rules (conditional probabilities) are used to set the strength of association of the attribute with a family ($s_{a,f}$). We use empirically determined Bayesian priors $p(f)$ to set the base-level of each family as opposed to using a constant base-level for instance based. Only two components of the activation function in Equation~\ref{actStr} are used, namely base-level and spreading activation. Given the attributes for current malware , we calculate the probability of the sample belonging to each family according to Equation~\ref{retprob}, generating a probability distribution over families. The task are then determined in a similar way to that of instance-based model. \smallskip

\algsetup{indent=2em}
\begin{algorithm}[]
	\caption{ACT-R Rule-based Learning}
	\begin{algorithmic}
		
		\STATE \textbf {INPUT:} New malware sample $i$, historical malware corpus $\M$.
		\STATE \textbf {OUTPUT:} Set of tasks associated with new sample $i$.

		\STATE \textbf{TRAINING:}
		\STATE Let $X=\bigcup_{j \in \M}attrib(j)$
		\FORALL {$a$ in $X$} 
		\STATE Compute the set of rules $p(a | f)$ and $p(a |\neg f)$\\ (where $p( a | f)=\frac{|\{i \in \M\cap f \ s.t.\ a \in attrib(i)\}|}{|f|}$\\ and $p( a | \neg f)=\frac{|\{i \in \M - f \ s.t. \ a \in attrib(i)\}|}{|\M|-|f|}$)
		\ENDFOR
		
		\STATE \textbf{TESTING:}
		
		\FORALL {$f\in \F$}
		\STATE $B_f = log(p(f))$ (where $p(f)=\frac{|f|}{|\M|}$)
		\STATE $s_{a,f}$ = 0.0
		\FORALL {$a \in attrib(i)$}
		\STATE $s_{a,f}$ = $log(\frac{p(a |f)}{p(a | \neg f)})$; $S_f$ =+ $\frac{w\times s_{a,f}}{|\attrib(i)|}$
		
		\ENDFOR
		
		\STATE $A_f$ = $B_f$ + $S_f$
		\ENDFOR
		\STATE Calculate $\Pr_f$ as per Equation~\ref{retprob}
		
		\STATE $t_p = \{t \in  T | p_f\geq0.5\}$

	\end{algorithmic}
\end{algorithm}

\subsection{Model Parameter Settings}
The two proposed models leverage separate components of the activation function.  Table~\ref{paramTable} provides a list of parameters used for both the ACT-R models - we use standard ACT-R parameters that have been estimated from a wide range of previous ACT-R modeling studies from other domains~\cite{parameters} and which are also suggested in the ACT-R reference manual~\cite{actr}. 

The intuition behind these parameters is as follows. The parameter $s$ injects stochastic noise in the model. It is used to compute the variance of the noise distribution and to compute the retrieval probability of each sample in memory. The mismatch penalty parameter $mp$ is an architectural parameter that is constant across samples, but it multiplies the similarity between the test sample and the samples in knowledge base.  Thus, with a large value it penalizes the mismatch samples more. It typically trades off against the value of the noise $s$ in a signal-to-noise ratio manner: larger values of $mp$ lead to more consistent retrieval of the closest matching sample whereas larger values of $s$ leads to more common retrieval of poorer matching samples.The activation threshold $\tau$ determines which samples will be retrieved from memory to make task prediction decisions. The base level constant $\beta$ is used to avoid retrieval failures which might be caused due to high activation threshold. The source activation $w$ is assigned to each retrieval to avoid retrieval failures for rule-based models.
\vspace{-1em}
\begin{table}[h!]
	\caption{\textmd{Parameters for the Cognitive models}}
	\label{paramTable}
	\centering
	\renewcommand{\arraystretch}{1.5}
	
	\begin{tabular}{|p{2.2cm}|p{5cm}|} 
		\hline
		{\bf Model} &  {\bf Parameters} \\ \hline 
		Instance Based Learning & $\beta$ = 20 (base-level constant)\newline $s$ = 0.1 (stochastic noise parameter)\newline $\tau$ = -10 (activation threshold)\newline $mp$ = 20(mismatch penalty) \\  \hline
		Rule Based learning & $s$ = 0.1 (stochastic noise parameter) \newline $w$ = 16 (source activation) \\ 
		
		\hline
	\end{tabular}
	\vspace{-1em}
	
\end{table}

\section{Experimental Setup}

\subsection{Baseline Approaches}
\label{baselineAppr}
We compare the proposed cognitive models against a variety of baseline approaches - one commercial package and five standard machine learning techniques.  For the machine learning techniques, we generate a probability distribution over families and return the set of tasks associated with a probability of $0.5$ or greater while the commercial software was used as intended by the manufacturer.  Parameters for all baseline approaches were set in a manner to provide the best performance.\smallskip\\
\noindent\textbf{Commercial Offering: Invencia Cynomix.}  Cynomix is a malware analysis tool made available to researchers by Invencia industries~\cite{invencia} originally developed under DARPA's Cyber Genome project.  It represents the current state-of-the-art in the field of malware capability detection.  Cynomix conducts static analysis of the malware sample and uses a proprietary algorithm to compare it to crowd-sourced identified malware components where the functionality is known.\smallskip\\
\noindent\textbf{Decision Tree (DT).}  Decision tree is a hierarchical recursive partitioning algorithm. We build the decision tree by finding the best split attribute i.e. the attribute that maximizes the information gain at each split of a node. In order to avoid over-fitting, the terminating criteria is set to less than 5\% of total samples. Malware samples are tested by the presence and absence of the best split attribute at each level in the tree till it reaches the leaf node.\smallskip\\
\noindent\textbf{Naive Bayes Classifier (NB).}  Naive Bayes is a probabilistic classifier which uses Bayes theorem with independent attribute assumption. During training we compute the conditional probabilities of a given attribute belonging to a particular family. We also compute the prior probabilities for each family i.e. fraction of the training data belonging to each family. Naive Bayes assumes that the attributes are statistically independent hence the likelihood for a sample $S$ represented with a set of attributes $a$ associated with a family $f$ is given as, \begin{math} \Pr(f | S) = P(f)\times \prod_{i=1}^{d} \Pr(a_i | f) \end{math}, where $d$ is the number of attributes in $a$.\smallskip\\
\noindent\textbf{Random Forest (RF).}  Ensemble methods are popular classification tools. It is based on the idea of generating multiple predictors used in combination to classify new unseen samples. We use a random forest which combines bagging for each tree with random feature selection at each node to split the data thus generating multiple decision tree classifiers~\cite{Breiman01}. Each decision tree gives its own opinion on test sample classification which are then merged to generate a probability distribution over families.\smallskip\\
\noindent\textbf{Support Vector Machine (SVM).}  Support vector machines (SVM) was proposed by Vapnik~\cite{Cortes95support-vectornetworks}. SVM's work by finding a separating margin that maximizes the geometric distance between classes. The separating margin is termed as hyperplane. We use the popular LibSVM implementation~\cite{Chang2011} which is publicly available.\smallskip\\
\noindent\textbf{Logistic Regression (LOG-REG).}  Logistic regression classifies samples by computing the odds ratio. The odds ratio gives the strength of association between the attributes and the family like simple rules used in the ACT-R rule based learning. We implement the multinomial logistic regression which handles multi-class classification.

\subsection{Dynamic Malware Analysis}
\label{sandbox}
Dynamic analysis studies a malicious program as it executes on the host machine. It uses tools like debuggers, function call tracers, machine emulators, logic analyzers, and network sniffers to capture the behavior of the program. 
We use two publicly available malware analysis tools to generate attributes for each malware sample. These tools make use of a sandbox which is a controlled environment to run malicious software.\smallskip\\
\noindent{\bf Anubis Sandbox.}  Anubis~\cite{anubis} is an online sandbox which generates an XML formated report for a malware execution in a remote environment. It generates detailed static analysis of the malware but provides less details regarding the behavior of the malware on the host machine. Since it is hosted remotely we cannot modify its settings.\smallskip\\
\noindent{\bf Cuckoo Sandbox.}  Cuckoo~\cite{cuckoo} is a standalone sandbox implemented using a dedicated virtual machine and more importantly can be customized to suit our needs. It generates detailed reports for both static as well as behavior analysis by watching and logging the malware while its running on the virtual machine. These behavior analysis prove to be unique indicators for a given malware for the experiments. 

\subsection{Performance Evaluation}
In our tests, we evaluate performance based primarily on four metrics: precision, recall, unbiased F1, and family prediction accuracy.  For a given malware sample being tested, precision is the fraction of tasks the algorithm associated with the malware that were actual tasks in the ground truth.  Recall, for a piece of malware, is the fraction of ground truth tasks identified by the algorithm.  The unbiased F1 is the harmonic mean of precision and recall.  In our results, we report the averages for precision, recall, and unbiased F1 for the number of trials performed.  Our measure of family accuracy - the fraction of trials where the most probable family was the ground truth family of the malware in question - is meant to give some insight into how the algorithm performs in the intermediate steps.

\section{Results}
\noindent All experiments were run on Intel core-i7 operating at 3.2 GHz with 16 GB RAM. Only one core was used for experiments. \textit{All experimental results presented in this section are new and have not been previously introduced.}
\label{expSec}
\subsection{Mandiant Dataset}
Our first set of experiments uses a dataset based on the the T1 cyber espionage group as identified in the popular report by Mandiant Inc~\cite{mandiant}.  This dataset consisted of $132$ real malware samples associated with the Mandiant report that were obtained from the Contagio security professional website~\cite{apt1sams}.  Each malware sample belonged to one of 15 families including BISCUIT, NEWSREELS, GREENCAT and COOKIEBAG. Based on the malware family description~\cite{mandiant}, we associated a set of tasks with each malware family (that each malware in that family was designed to perform). In total, 30 malware tasks were identified for the given malware samples (see Table~\ref{exTask}). On average, each family performed 9 tasks.

\begin{figure}[htp!]
	\centerline{\includegraphics[scale=0.25,keepaspectratio]{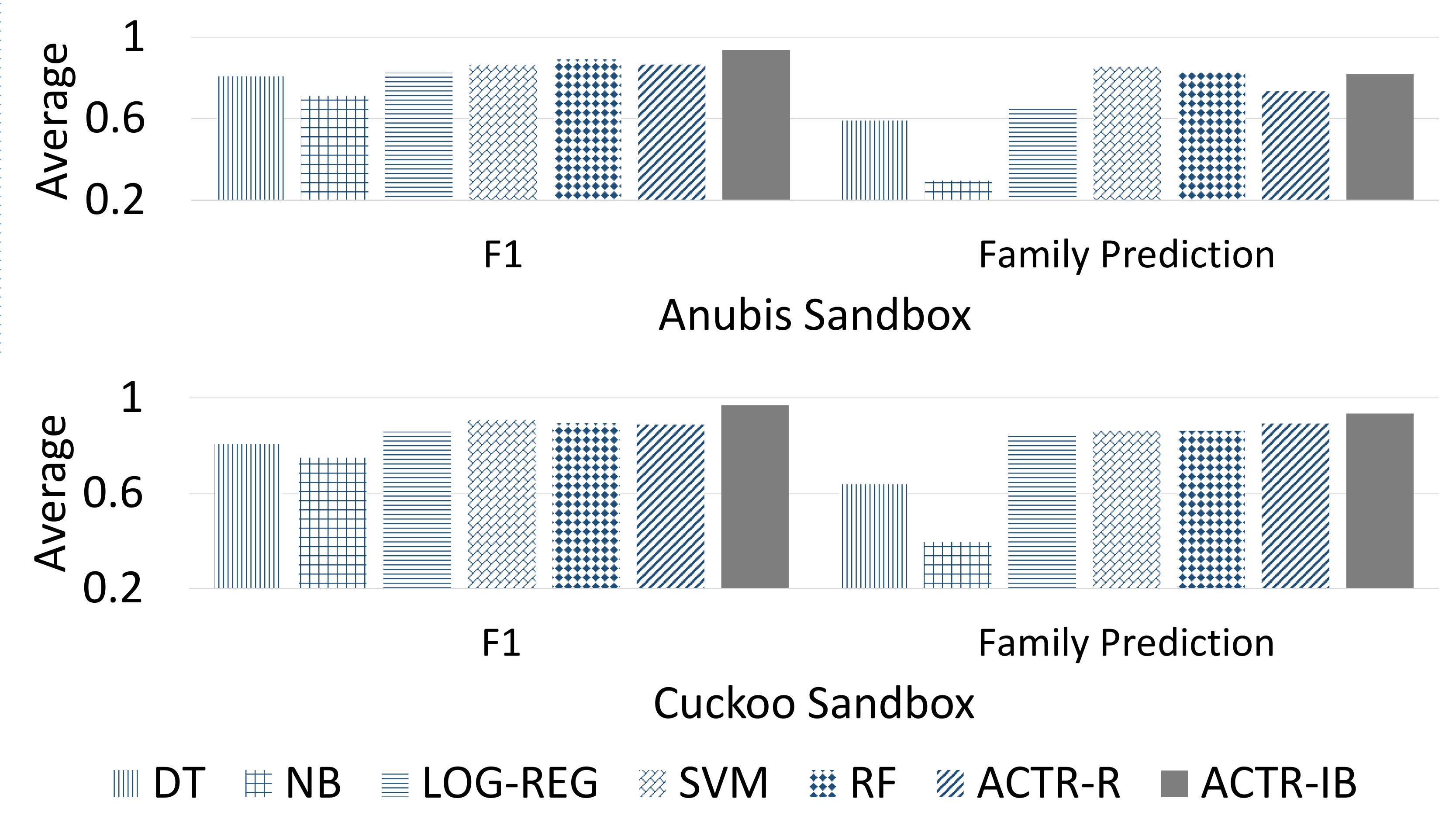}}
	\vspace{0em}
	\caption{\textmd{Average F1 and Family prediction comparisons for DT, NB, LOG-REG, SVM, RF, ACTR-IB and ACTR-R for Anubis (top) and Cuckoo (bottom).}}
	\vspace{-1em}
	\label{fig:one}
\end{figure}

We compared the four machine learning approaches with the rule based and instance-based ACT-R models (ACTR-R and ACTR-IB respectively). We also submitted the samples to the Cynomix tool for automatic detection of capabilities. These detected capabilities were then manually mapped to the tasks from the Mandiant report. Precision and recall values were computed for the inferred adversarial tasks. On average the machine learning approaches predicted 9 tasks per sample, ACTR-R predicted 9 tasks per sample and ACTR-IB predicted 10 tasks. On the other hand Cynomix was able to detect on average only 4 tasks.\\\\
{\bf Leave one out Cross-Validation(LOOCV)}\\
In leave one out cross validation, for $n$ malware samples, we train on $n-1$ samples and test on the remaining one. This procedure was repeated for all samples and the results were averaged. We performed this experiment using both sandboxes and compared the results (see \figurename~\ref{fig:one}).

The average F1 increases by 0.03 when we use the attributes generated by the Cuckoo sandbox instead of Anubis. The statistical significance results are as follows: for ACTR-IB  (t (132) = 1.94, p = 0.05), ACTR-R (t (132) = 1.39, p = 0.16), RF (t (132) = 0.56, p = 0.57), SVM (t (132) = 1.95, p = 0.05), LOG-REG (t (132) = 1.82, p = 0.07), NB (t (132) = 1.79, p = 0.08) and DT (t (132) = 0.83, p = 0.4). But the significant improvement was in the family prediction values with ACTR-IB improving by 0.12 from 0.81 to 0.93 (t (132) = 3.86, p $<$ .001) and ACTR-R by 0.15 from 0.72 to 0.87 (t (132) = 3.78, p $<$ .001) outperforming all other methods. Since having behavior analysis helps in better task prediction as seen from the comparison experiment, we use cuckoo sandbox for rest of our experiments.

\begin{figure}[htp!]
	\centerline{\includegraphics[scale=0.22,keepaspectratio]{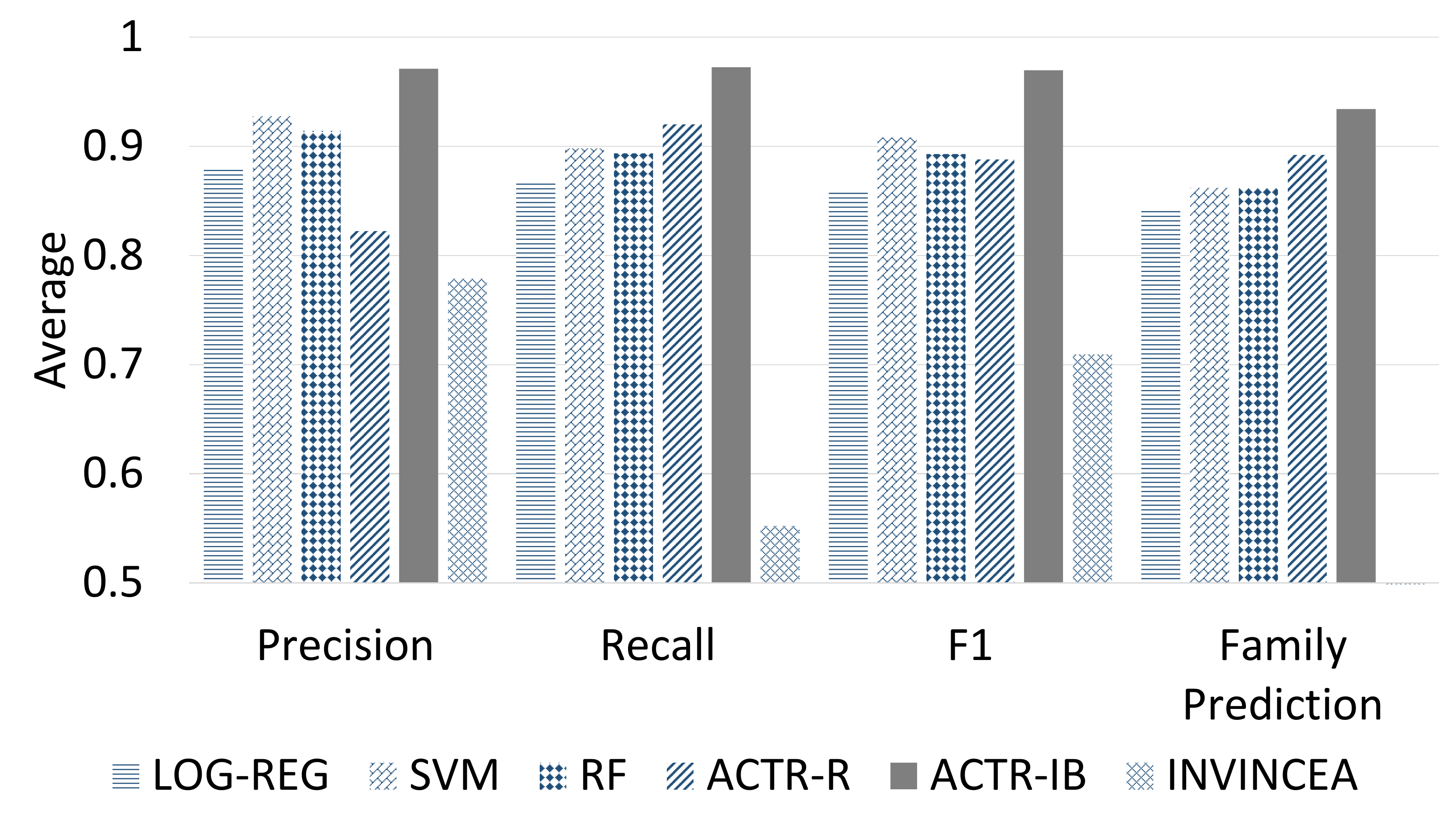}}
	\vspace{0em}
	\caption{\textmd{Average Precision, Recall, F1 and Family prediction comparisons for LOG-REG, RF, SVM,  ACTR-R, ACTR-IB and INVINCEA.}}
	\vspace{-1em}
	\label{fig:two}
\end{figure}
\figurename~\ref{fig:two} compares the performance of the five best performing methods from \figurename~\ref{fig:one} and compares it with the Cynomix tool of Invincea industries. ACTR-IB outperformed LOG-REG, SVM, RF and ACTR-R; average F1 = 0.97 vs 0.85 (t (132) = 7.85,   p $<$ .001), 0.9 (t (132) = 4.7, p $<$ .001),  0.89 (t (132) = 5.45,   p $<$ .001) and 0.88 (t (132) = 5.2, p $<$ .001) respectively. Both the proposed cognitive models and machine learning techniques significantly outperformed the Cynomix tool in detecting the capabilities (tasks).

These three approaches (LOG-REG, SVM, RF) were also evaluated with respect to predicting the correct family (before the tasks were determined). ACTR-IB outperformed LOG-REG, SVM, RF and ACTR-R; average family prediction = 0.93 vs 0.84 (t (132) = 3.22,   p $<$ .001), 0.86 (t (132) = 3.13, p $<$ .001),  0.86 (t (132) = 3.13,   p $<$ .001) and 0.89 (t (132) = 2.13, p = .03) respectively.\\\\
{\bf Task Prediction without inferring families:}\\
In the proposed models we infer the malware family first and then predict the tasks associated with that family. However, differences over ``ground truth'' for malware families in the cyber-security community calls for a direct inference of tasks without dependence on family prediction. In this section we adapt the models to predict tasks directly without inferring the family. 
\begin{figure}[htp!]
	\centerline{\includegraphics[scale=0.2,keepaspectratio]{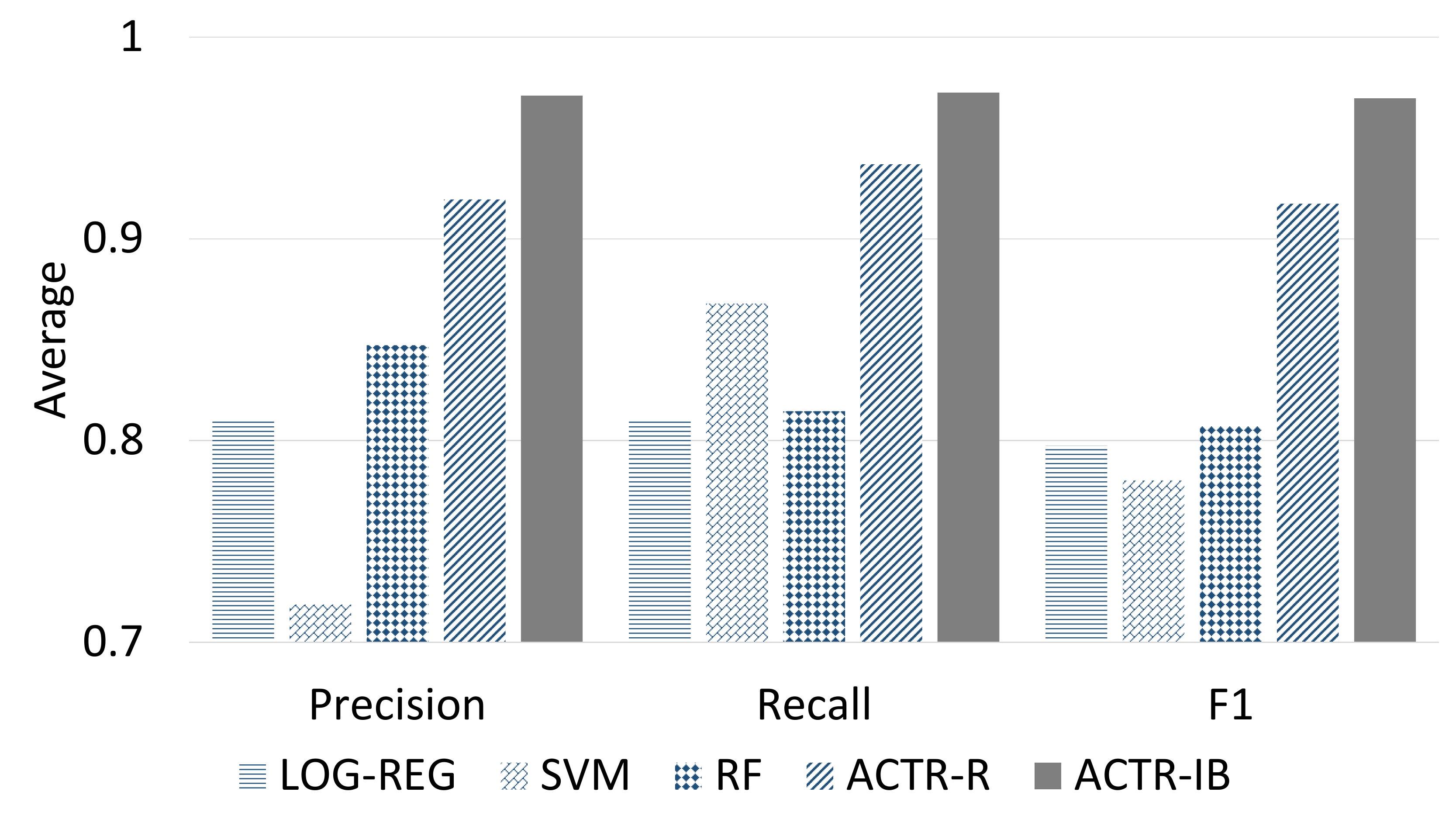}}
	\vspace{0em}
	\caption{\textmd{Average Precision, Recall, and F1 comparisons for LOG-REG, RF, SVM,  ACTR-R and ACTR-IB without inferring families.}}
	\vspace{-1em}
	\label{fig:nf}
\end{figure}

\begin{figure}[ht]
	
	\centering
	\begin{minipage}[t]{0.47\linewidth}
	
		\includegraphics[width=\textwidth,keepaspectratio]{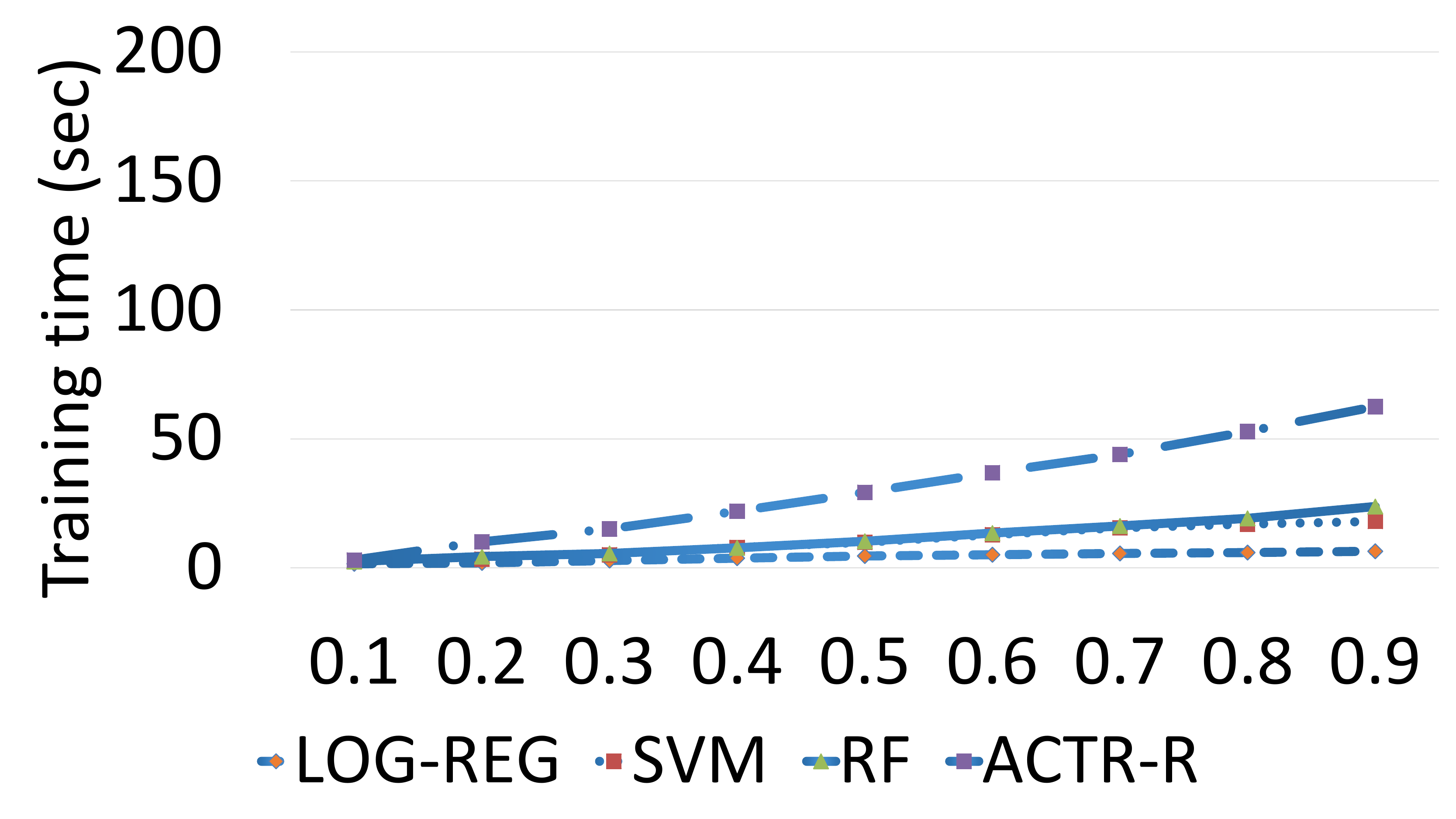}
		\subcaption{(a)}
		\label{fig:minipage1}
	\end{minipage}
	\quad
	\begin{minipage}[t]{0.47\linewidth}
		
		\includegraphics[width=\textwidth, keepaspectratio]{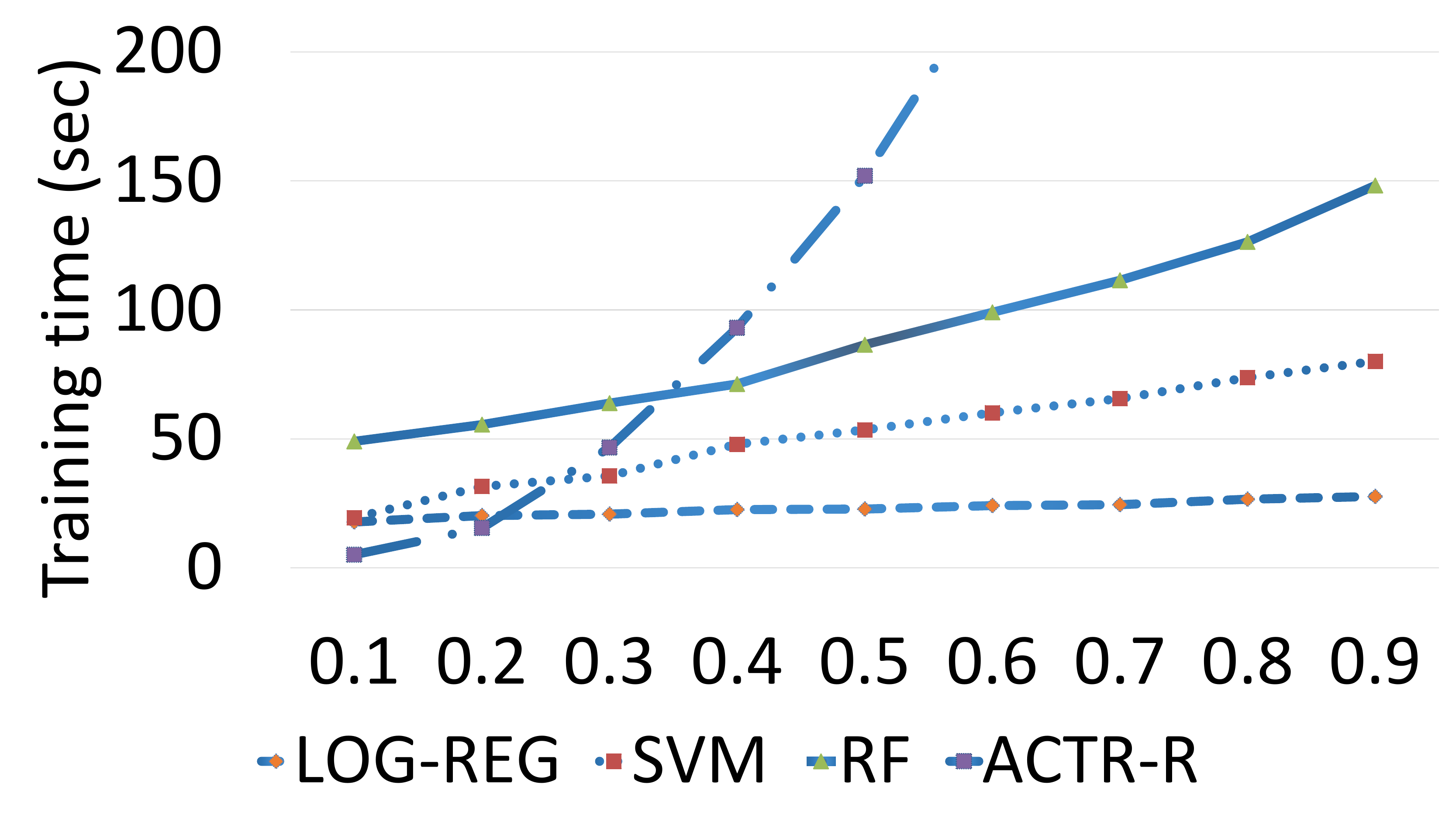}
		\subcaption{(b)}
		\label{fig:minipage2}
	\end{minipage}
	\vspace{-1em}
	\caption{\textmd{Training time for LOG-REG, SVM, RF and ACTR-R with(a) / without(b) inferring families.}}
	\label{t}
\end{figure}

\figurename~\ref{fig:nf} shows the performance of the cognitive and machine learning models without inferring the families. There is no difference in the performance of ACTR-IB and ACTR-R approaches as compared to \figurename~\ref{fig:two} where we use families.  On the other hand direct task prediction reduces the F1 measure of machine learning techniques on average by almost 0.1. This is due to the fact, now instead of having a single classifier for each family we have multiple classifiers for each task that a malware sample is designed to perform. This not only degrades the performance but also adds to the training time for these methods. We compare the training time with increase in training data for task prediction with/without inferring families. Inferring families first reduces the training time (see \figurename~\ref{t} (a)).  On the other hand predicting tasks directly significantly increases the training time for the machine learning methods along for the rule-based ACT-R approach (\figurename~\ref{t} (b)). Due to the issues with respect to performance and training time, we consider inferring families first for rest of the experiments. An important point to note is this has no effect on the Instance-based model for both performance and computation time.\\\\
\vspace{-2em}
\vspace{-2em}

\subsection{GVDG Dataset}

\begin{figure}[htb!]
	\centerline{\includegraphics[scale=0.38,keepaspectratio]{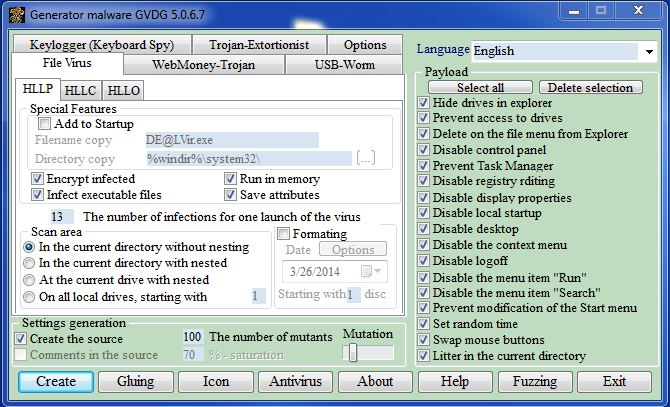}}
	
	\caption{\textmd{GVDG User Interface}}
	
	\label{gvdg}
\end{figure}
\vspace{-2em}
GVDG is a malware generation tool designed for the study of computer threats~\cite{gvdg}. It is capable of generating following malware threats,
\begin{itemize}
	\setlength\itemsep{-0.1em}
	\item File-virus
	\item Key-Logger
	\item Trojan-Extortionist 
	\item USB-Worm
	\item Web Money-Trojan
\end{itemize}

\begin{figure}[htb!]
	\centerline{\includegraphics[scale=0.2,keepaspectratio]{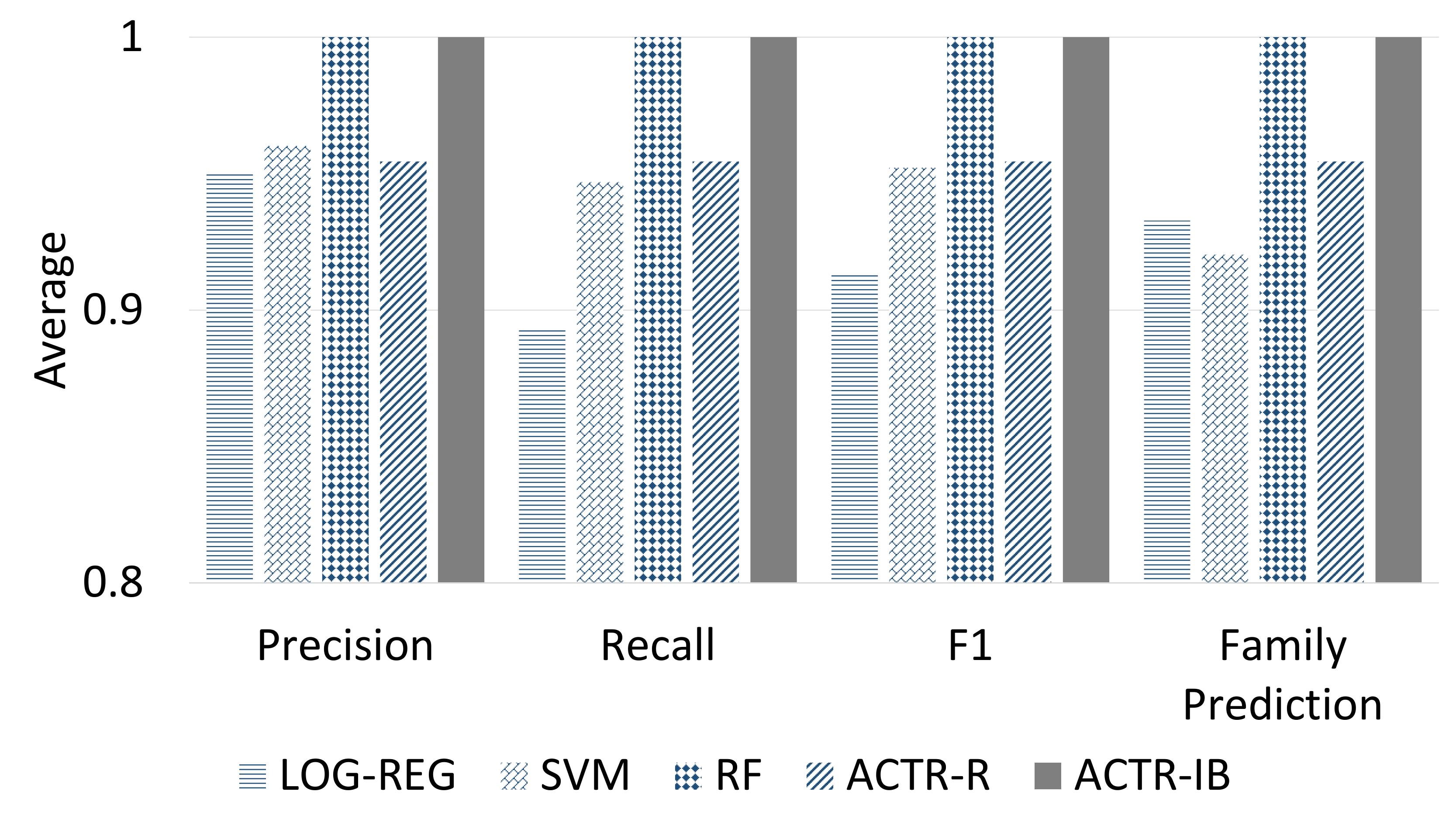}}
	\vspace{0em}
	\caption{\textmd{Average Precision, Recall, F1 and Family prediction comparisons for LOG-REG,SVM, RF, ACTR-R and ACTR-IB for different carrier samples.}}
	\vspace{-1em}
	\label{gvdg:1}
\end{figure}

\figurename~\ref{gvdg} shows the GVDG user interface used for the generation of malware samples. We can select the carrier type and the tasks that we want the malware sample to perform on the host machine. The tasks are represented as payloads, while carrier is a functional template of specific behavior which are the operational framework supporting and enabling the task activity. In generating datasets with GVDG, we specify families based on sets of malware with the same tasks.  Whether or not a family consists of malware with the same carrier depends on the experiment.  Further, GVDG also has an option to increase ``mutation'' or variance among the samples. We perform experiments analyzing the performance of the proposed methods when the generated samples belong to different carrier and same carrier types, as well as when the samples are encrypted and mutated making task prediction difficult.  In all the experiments we consider 60\% of the data for training and 40\% for testing. The results are averaged across 10 trials. The Cynomix tool from Invencia was unable to detect any tasks for the GVDG dataset, primarily due to to its inability to find public source documents referencing GVDG samples and also unable to generalize from similar samples.\\\\
{\bf Different Carriers:}\\
In this experiment, we generated 1000 samples for each carrier type with low mutation. On average each carrier type performs 7 tasks(payloads). Hence each carrier represents one family for this experiment. Both random forest and ACTR-IB model were able to predict the tasks and family with F1 measure of 1.0 outperforming LOG-REG 1 vs 0.91 , SVM 1 vs 0.95 and ACTR-R 1 vs 0.95. All results are statistical significant with (t (1998) $\geq$ 8.93, p $<$ .001)(\figurename~\ref{gvdg:1}). Also for family prediction ACTR-IB and RF outperformed LOG-REG 1 vs 0.92, SVM 1 vs 0.92  and ACTR-R 1 vs 0.95 (t (1998) $\geq$ 8.93, $<$ .001).  These results are not surprising given that different carrier(family) types have high dissimilarity between them. Also, samples belonging to the same carrier have on average 60\% of similar attributes.

\begin{figure}[htb!]
	\centerline{\includegraphics[scale=0.2,keepaspectratio]{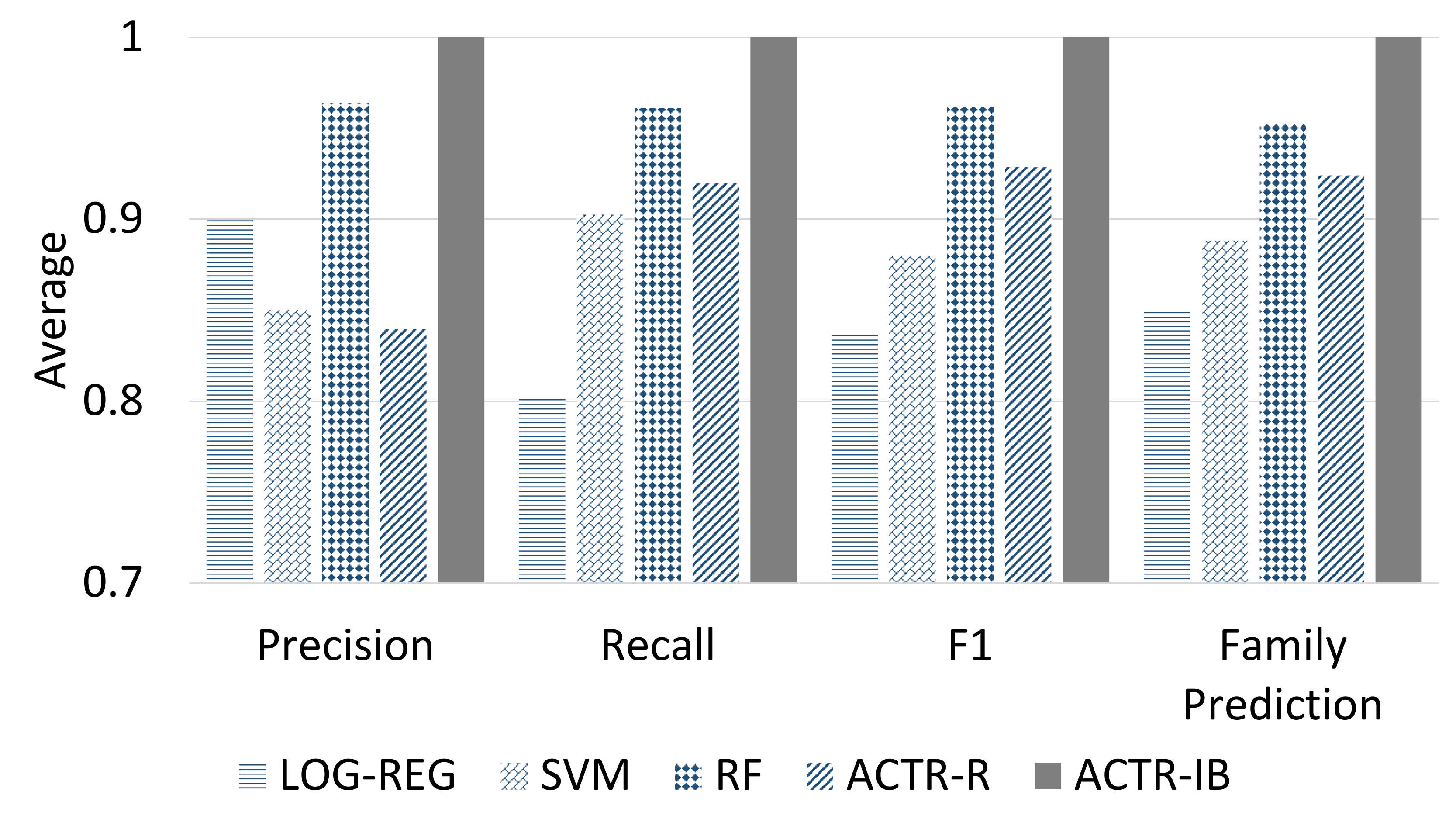}}
	\vspace{0em}
	\caption{\textmd{Average Precision, Recall, F1 and Family prediction comparisons for LOG-REG,SVM, RF, ACTR-R and ACTR-IB.}}
	\vspace{-2em}
	\label{gvdg:2}
\end{figure}

\noindent
{\bf Different Carriers-Mutation:}\\
For this case, we generate the same samples as in the previous experiment but with maximum mutation between samples belonging to the same carrier. We generated 1000 samples for each carrier with maximum mutation. In this case ACTR-IB had an average F1 of 1 outperforming LOG-REG 1 vs 0.83, SVM 1 vs 0.88  , RF 1 vs 0.96 and ACTR-R 1 vs 0.92 (t (1998) $\geq$ 7, p $<$ .001)(\figurename~\ref{gvdg:2}). Also for family prediction ACTR-IB outperformed LOG-REG 1 vs 0.85, SVM 1 vs 0.88  , RF 1 vs 0.95 and ACTR-R 1 vs 0.92 (t (1998) $\geq$ 7, p $<$ .001). 

High mutation induces high variance between samples associated with the same carrier making the classification task difficult. High mutation samples belonging to same carrier have only 20\% of common attributes as compared to 60\% for low mutation.

\begin{figure}[htb!]
	\centerline{\includegraphics[scale=0.24,keepaspectratio]{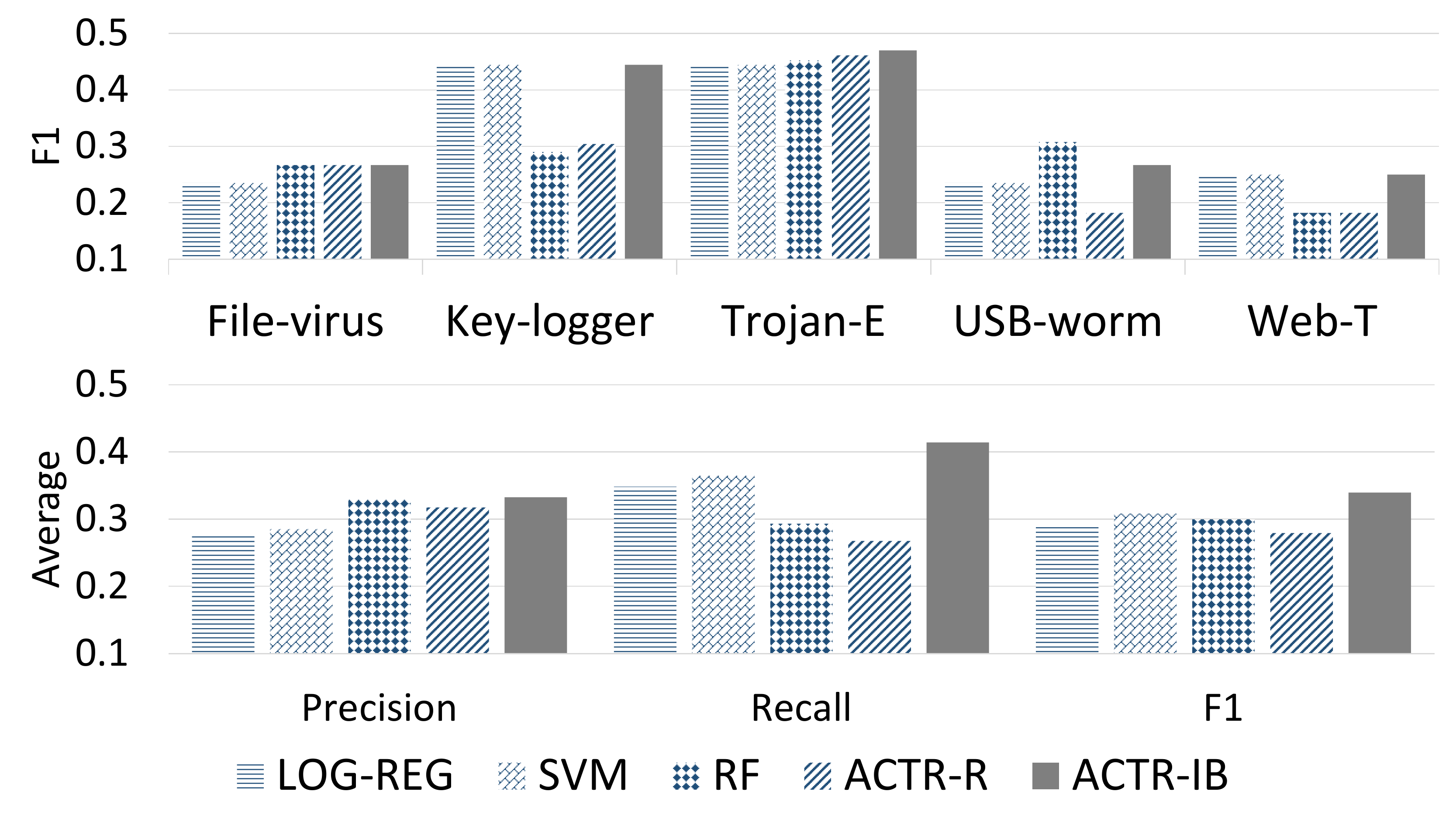}}
	\vspace{0em}
	\caption{\textmd{Average F1 values for 5 malware carriers (above) and the average precision, recall and F1 across all carriers (below) for LOG-REG, SVM, RF, ACTR-R and ACTR-IB..}}

	\label{gvdg:3}
\end{figure}

\noindent
{\bf Leave one carrier out cross-validation:}\\
To see how the models generalize to unseen malware family(carrier), we performed a leave-one-carrier-out comparison (using high mutation samples), where we test the models against one previously unseen malware carrier.  ACTR-IB performs better or on par with all other baseline approaches for all the carriers.  It clearly outperforms all the approaches in recalling most of the actual tasks (40\%) (see Figure~\ref{gvdg:3}). ACTR-IB has shown to generalize for unseen malware families~\cite{iccm:2015}. This case is difficult given the fact that the test family is not represented during training, hence task prediction depends on associating the test family with the training families that perform similar tasks. 

\noindent
{\bf Same Carrier:}\\
\vspace{-1em}
\begin{figure}[htb!]
	\centerline{\includegraphics[scale=0.2,keepaspectratio]{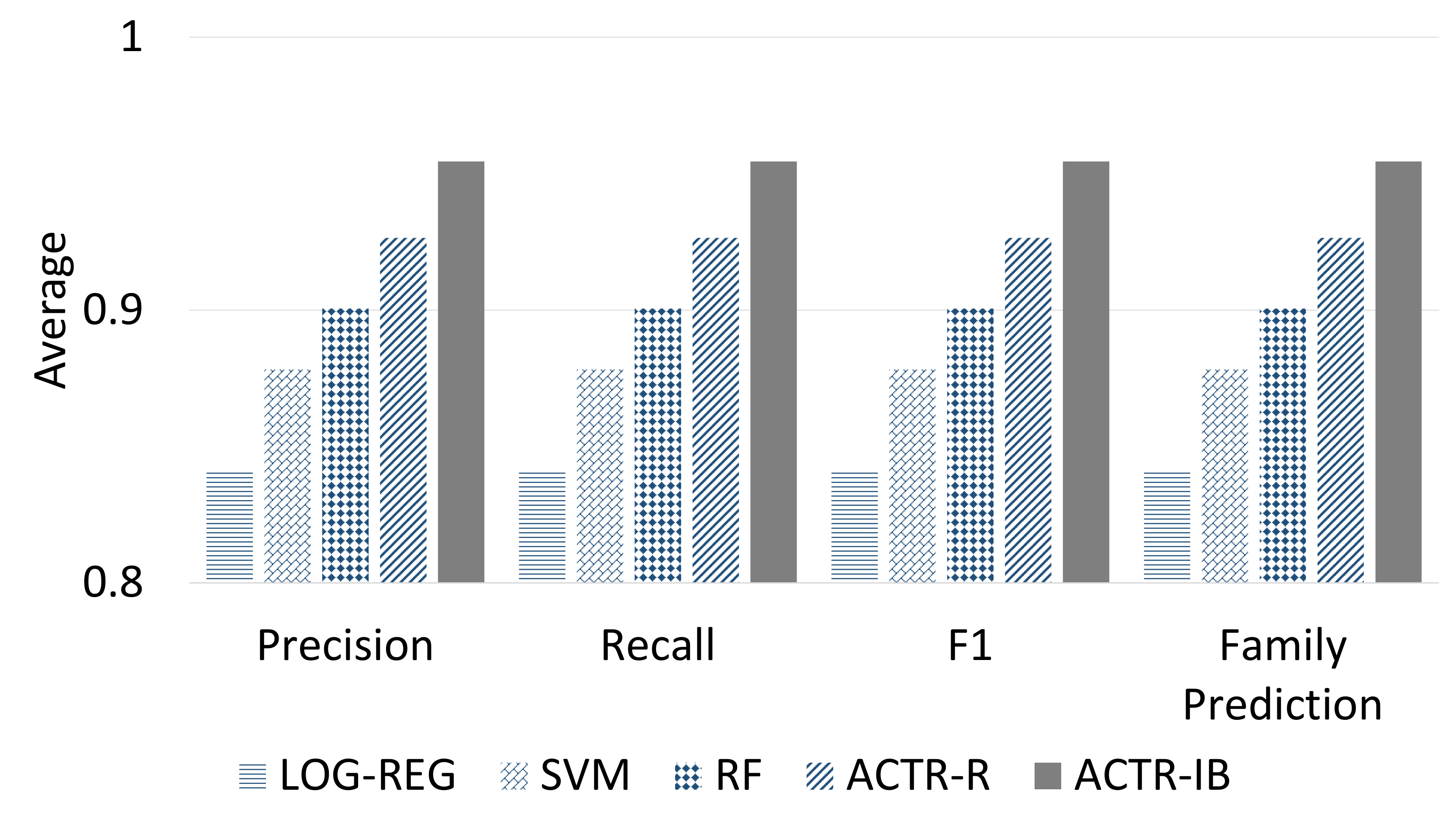}}
	\vspace{0em}
	\caption{\textmd{Average Precision, Recall, F1 and Family prediction comparisons for LOG-REG,SVM, RF, ACTR-R and ACTR-IB.}}
	\vspace{-2em}
	\label{gvdg:4}
\end{figure}

As seen in the previous experiments, different carrier types makes the task easier because of less similarity between them. We now test the performance, on same carrier type performing exactly one task. Since there are 17 tasks in the GVDG tool, we generate 100 samples for each task for carrier type File-virus. In this experiment each task represents one family. Thus in total we have 1700 samples. We do the 60-40 split experiment. From \figurename~\ref{gvdg:4}  ACTR-IB had an average F1 of 0.95 outperforming LOG-REG 0.95 vs 0.84, SVM 0.95 vs 0.87, RF 0.95 vs 0.90 and ACTR-R 0.95 vs 0.92 (t (678) $\geq$ 1.52 , p $\leq$ 0.13). Since each family performs exactly one task the family prediction is similar to F1. Using the same carrier for each payload makes the task difficult as they have high similarity between them. 

\begin{figure}[htb!]
	\centerline{\includegraphics[scale=0.2,keepaspectratio]{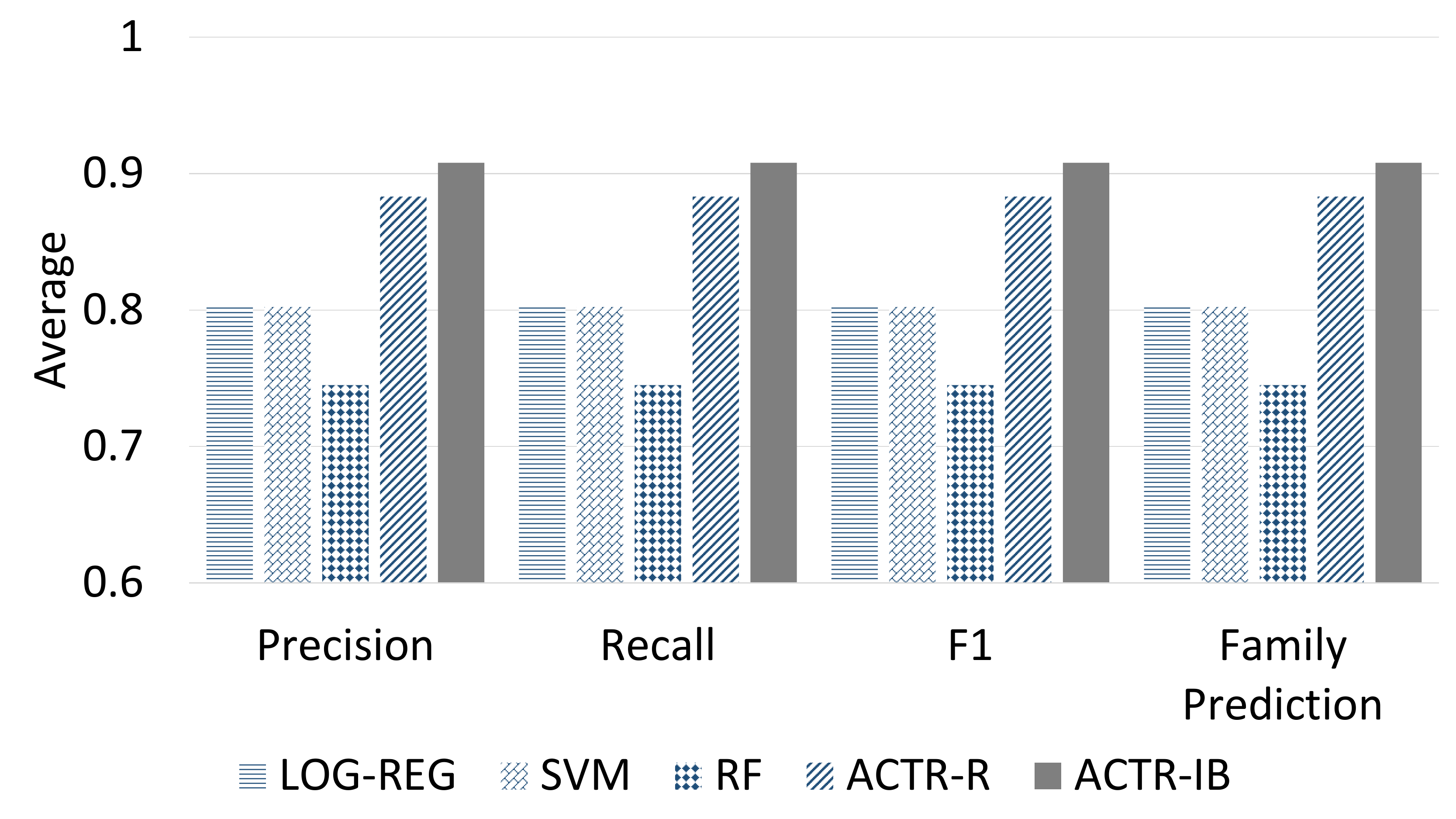}}
	\vspace{-1em}
	\caption{\textmd{Average Precision, Recall, F1 and Family prediction comparisons for LOG-REG,SVM, RF, ACTR-R and ACTR-IB.}}
	\vspace{-2.5em}
	\label{gvdg:5}
\end{figure}

\noindent
{\bf Same Carrier-Encryption:}\\	
The GVDG tool provides the option for encrypting the malware samples for the File-virus carrier type. We use this option to generate 100 encrypted malware samples for each task(payload) and use them as test data with the unencrypted versions from the same carrier experiment as training samples. From \figurename~\ref{gvdg:5} ACTR-IB had an average F1 of 0.9 outperforming LOG-REG 0.9 vs 0.8, SVM 0.9 vs 0.8, RF 0.9 vs 0.74 and ACTR-R 0.9 vs 0.88 (t (1698) $\geq$ 2.36  , p $\leq$ 0.02). Encrypting malware samples morphs the task during execution making it difficult to detect during analysis. Hence the drop in performance as compared to non-encrypted samples. We note that SVM performs better than RF  likely because it looks to maximize generalization.\\

\section{Related Work}
\label{rwSec}
\noindent\textbf{Identification of malicious software.}  The identification of whether or not binary is malicious~\cite{Firdausi:2010, Tamersoy:2014} is important and can be regarded as a ``first step'' in the analysis of suspicious binaries in the aftermath of a cyber-attack.  However, we note that as many pieces of malware are designed to perform multiple tasks, that successful identification of a binary as malicious does not mean that the identification of its associated tasks will be a byproduct of the result - and hence this is normally the case, which has led to some of the other related work described in this section.\smallskip\\
\noindent\textbf{Malware family classification.}  There is a wealth of existing work on malware family identification~\cite{Bayer_scalable, Kinable:2011, Kong:2013}.  The intuition here is that by identifying the family of a given piece of malware, an analyst can then more easily determine what it was designed to do based on previously studied samples from the same family. However, malware family classification has suffered from two primary draw-backs: (1) disagreement about malware family ground truth as different analysts (e.g. Symantec and MacAfee) cluster malware into families differently; and (2) previous work has shown that some of these approaches mainly succeed in ``easy to classify'' samples ~\cite{Li07onchallenges,PerdisciU12}, where ``easy to classify'' is a family that is agreed upon by multiple malware firms.  In this paper, we infer the specific tasks a piece of malware was designed to carry out.  While we do assign malware to a family as a component of our approach, to avoid the two aforementioned issues as the family partition is done so probabilistically and the result ground truth is the focus of our comparison (though we show family prediction results as a side-result).  Further, we also describe and evaluate a variant of our instance-based method that does not consider families and yields a comparable performance to our instance-based method that does consider families.
\smallskip\\
\noindent\textbf{Malware task identification.}  With regard to direct inference of malware tasks, the major related work include the software created by the firm Invincea~\cite{invencia} for which we have included a performance comparison.  Additionally, some of the ideas in this paper were first introduced in~\cite{iccm:2015,brims}.  However, that work primarily focused on describing the intuitions behind the cognitive modeling techniques and only included experimental evaluation on one dataset (the Mandiant APT1 dataset) and one sandbox environment (Anubis) with a comparison amongst only the instance based approach, the rule-based cognitive model, the standard decision tree, and the naive Bayes reasoner.  The experimental evaluation in this paper was designed to be much more thorough to pave the way toward deployment of the approach for use by cyber-security analysts.

\section{Conclusion}
\label{con}
In this paper, we introduced an automated method that combines dynamic malware analysis with cognitive modeling to identify malware tasks.  This method obtains excellent precision and recall - often achieving an unbiased F1 score of over 0.9 - in a wide variety of conditions over two different malware sample collections and two different sandbox environments - outperforming a variety of baseline methods.
Currently, our future work has three directions.  First, we are looking to create a deployed version of our approach to aid cyber-security analysts in the field.  Second, we look to enhance our malware analysis to also include network traffic resulting from the sample by extending the capabilities of the sandbox.  Finally, we also look to address cases of highly-sophisticated malware that in addition to using encryption and packing to limit static analysis, also employ methods to ``shut down'' when run in a sandbox environment~\cite{Lindorfer:2011}.  We are exploring multiple methods to address this such as the recently introduced technique of ``spatial analysis''~\cite{sentar} that involves direct analysis of a malware binary.

\bibliographystyle{abbrv}

\bibliography{malwareTaskIdBib}

\begin{thebibliography}{10}

\bibitem{Anderson04anintegrated}
J.~R. Anderson, D.~Bothell, M.~D. Byrne, S.~Douglass, C.~Lebiere, and Y.~Qin.
\newblock An integrated theory of mind.
\newblock {\em PSYCHOLOGICAL REVIEW}, 111:1036--1060, 2004.

\bibitem{Bayer_scalable}
U.~Bayer, P.~M. Comparetti, C.~Hlauschek, C.~Kruegel, and E.~Kirda.
\newblock Scalable, behavior-based malware clustering, 2009.

\bibitem{actr}
D.~Bothell.
\newblock Act-r 6.0 reference manual.
\newblock {\em \url{http://act-r.psy.cmu.edu/actr6/reference-manual.pdf}},
  2004.

\bibitem{Breiman01}
L.~Breiman.
\newblock Random forests.
\newblock {\em Machine Learning}, 45(1):5--32, 2001.

\bibitem{Chang2011}
C.-C. Chang and C.-J. Lin.
\newblock Libsvm: A library for support vector machines.
\newblock {\em ACM Trans. Intell. Syst. Technol.}, 2(3):27:1--27:27, May 2011.

\bibitem{cuckoo}
J.~B. M.~S. Claudio~Guarnieri, Alessandro~Tanasi.
\newblock Cuckoo sandbox.
\newblock {\em \url{http://www.cuckoosandbox.org/}}, 2012.

\bibitem{Cortes95support-vectornetworks}
C.~Cortes and V.~Vapnik.
\newblock Support-vector networks.
\newblock pages 273--297, 1995.

\bibitem{Firdausi:2010}
I.~Firdausi, C.~lim, A.~Erwin, and A.~S. Nugroho.
\newblock Analysis of machine learning techniques used in behavior-based
  malware detection.
\newblock In {\em Proceedings of the 2010 Second International Conference on
  ACT}, ACT '10, pages 201--203, Washington, DC, USA, 2010. IEEE Computer
  Society.

\bibitem{sentar}
D.~Giametta and A.~Potter.
\newblock Shmoomcon 2014:there and back again:a critical analysis of spatial
  analysis, 2014.

\bibitem{Gonzalez2003591}
C.~Gonzalez, J.~F. Lerch, and C.~Lebiere.
\newblock Instance-based learning in dynamic decision making.
\newblock {\em Cognitive Science}, 27(4):591 -- 635, 2003.

\bibitem{gvdg}
GVDG.
\newblock Generator malware gvdg.
\newblock 2011.

\bibitem{invencia}
Invencia.
\newblock Crowdsource: Crowd trained machine learning model for malware
  capability detection.
\newblock {\em \url{http://www.invincea.com/tag/cynomix/}}, 2013.

\bibitem{anubis}
ISEC-Lab.
\newblock Anubis: Analyzing unknown binaries.
\newblock {\em \url{http://anubis.iseclab.org/}}, 2007.

\bibitem{Gauss}
Kaspersky.
\newblock Gauss: Abnormal distribution, 2012.

\bibitem{Kinable:2011}
J.~Kinable and O.~Kostakis.
\newblock Malware classification based on call graph clustering.
\newblock {\em J. Comput. Virol.}, 7(4):233--245, Nov. 2011.

\bibitem{Kong:2013}
D.~Kong and G.~Yan.
\newblock Discriminant malware distance learning on structural information for
  automated malware classification.
\newblock In {\em Proceedings of the 19th ACM SIGKDD}, KDD '13, pages
  1357--1365, New York, NY, USA, 2013. ACM.

\bibitem{iccm:2015}
C.~Lebiere, S.~Bennati, R.~Thomson, P.~Shakarian, and E.~Nunes.
\newblock Functional cognitive models of malware identification.
\newblock In {\em Proceedings of ICCM, {ICCM} 2015, Groningen, The Netherlands,
  April 9-11, 2015}, 2015.

\bibitem{Lebiere:2013}
C.~Lebiere, P.~Pirolli, R.~Thomson, J.~Paik, M.~Rutledge-Taylor, J.~Staszewski,
  and J.~R. Anderson.
\newblock A functional model of sensemaking in a neurocognitive architecture.
\newblock {\em Intell. Neuroscience}, 2013:5:5--5:5, Jan. 2013.

\bibitem{Li07onchallenges}
P.~Li, L.~Liu, and M.~K. Reiter.
\newblock On challenges in evaluating malware clustering, 2007.

\bibitem{Lindorfer:2011}
M.~Lindorfer, C.~Kolbitsch, and P.~Milani~Comparetti.
\newblock Detecting environment-sensitive malware.
\newblock In {\em Proceedings of the 14th International Conference on RAID},
  RAID'11, pages 338--357, Berlin, Heidelberg, 2011. Springer-Verlag.

\bibitem{mandiant}
Mandiant.
\newblock Apt1:exposing one of china's cyber espionage units.
\newblock {\em \url{http://intelreport.mandiant.com/}}, 2013.

\bibitem{apt1sams}
Mandiant.
\newblock {Mandiant APT1 samples categorized by malware families}.
\newblock {\em {Contagio Malware Dump}}, 2013.

\bibitem{PerdisciU12}
R.~Perdisci and ManChon.
\newblock Vamo: towards a fully automated malware clustering validity analysis.
\newblock In {\em ACSAC}, pages 329--338. ACM, 2012.

\bibitem{sikorski}
M.~Sikorski and A.~Honig.
\newblock {\em Practical Malware Analysis: The Hands-On Guide to Dissecting
  Malicious Software}.
\newblock No Starch Press, 1 edition, 2012.

\bibitem{Sutton:1998}
R.~S. Sutton and A.~G. Barto.
\newblock {\em Introduction to Reinforcement Learning}.
\newblock MIT Press, Cambridge, MA, USA, 1st edition, 1998.

\bibitem{Tamersoy:2014}
A.~Tamersoy, K.~Roundy, and D.~H. Chau.
\newblock Guilt by association: Large scale malware detection by mining
  file-relation graphs.
\newblock In {\em Proceedings of the 20th ACM SIGKDD}, KDD '14, pages
  1524--1533. ACM, 2014.

\bibitem{brims}
R.~Thomson, C.~Lebiere, S.~Bennati, P.~Shakarian, and E.~Nunes.
\newblock Malware identification using cognitively-inspired inference.
\newblock In {\em Proceedings of BRIMS, {BRIMS} 2015, Washington DC, March
  31-April 3, 2015}, 2015.

\bibitem{parameters}
T.~J. Wong, E.~T. Cokely, and L.~J. Schooler.
\newblock An online database of act-r parameters: Towards a transparent
  community-based approach to model development.
\newblock 2010.

\end{thebibliography}

\end{document}